\documentclass[useAMS,usenatbib]{mn2e}

\usepackage{times}  
\usepackage{listings}
\usepackage{graphics}
\usepackage{subfigure}
\usepackage{graphicx}
\usepackage{graphicx,xspace}
\usepackage{amssymb}
\usepackage{hyperref}
\usepackage{aas_macros}
\usepackage{lscape}
\usepackage{rotating}
\usepackage{placeins}
\usepackage{natbib}
\usepackage{amsmath} 
\usepackage{ulem} 

\def\lsim{\mathrel{\hbox{\rlap{\hbox{\lower4pt\hbox{$\sim$}}}\hbox{$<$}}}}
\def\gsim{\mathrel{\hbox{\rlap{\hbox{\lower4pt\hbox{$\sim$}}}\hbox{$>$}}}}

\newcommand{\galform}{{\sc{galform}}\xspace}
\newcommand{\vcut}{$v_{\mathrm{cut}}$\xspace}
\newcommand{\zcut}{$z_{\mathrm{cut}}$\xspace}
\newcommand{\ahot}{$\alpha_{\mathrm{hot}}$\xspace}
\newcommand{\msun}{$\mathrm{M}_{\odot}$\xspace}
\newcommand{\mwdm}{$m_{\mathrm{WDM}}$\xspace}

\title{Constraining the WDM Particle Mass with Milky Way Satellites}
\author[Rachel Kennedy]{Rachel Kennedy$^{1}$\thanks{E-mail:
rachel.kennedy@durham.ac.uk}, Carlos Frenk$^{1}$, Shaun Cole$^{1}$, 
Andrew Benson$^{2}$\\
$^{1}$Institute for Computational Cosmology, Department of Physics, University of Durham, South Road, Durham, DH1 3LE, UK\\
$^{2}$Carnegie Observatories, 813 Santa Barbara Street, Pasadena, CA 91101, USA}

\begin{document}
\date{\today}
\pagerange{\pageref{firstpage}--\pageref{lastpage}} \pubyear{2013}
\maketitle
\label{firstpage}


\begin{abstract}
  Well-motivated particle physics theories predict the existence of
  particles (such as sterile neutrinos) which acquire non-negligible
  thermal velocities in the early universe. These particles could
  behave as warm dark matter (WDM) and generate a small-scale cutoff
  in the linear density power spectrum which scales approximately
  inversely with the particle mass. If this mass is of order a keV,
  the cutoff occurs on the scale of dwarf galaxies. Thus, in WDM
  models the abundance of small galaxies, such as the satellites that
  orbit in the halo of the Milky Way, depends on the mass of the warm
  particle. The abundance also scales with the mass of the host
  galactic halo. We use the \galform semi-analytic model of galaxy
  formation to calculate the properties of galaxies in universes in
  which the dark matter is warm. Using this method, we can compare the
  predicted satellite luminosity functions to the observed data for
  the Milky Way dwarf spheroidals, and determine a lower bound on the
  thermally produced WDM particle mass. This depends strongly on the value of the Milky
  Way halo mass and, to some extent, on the baryonic physics assumed;
  we examine both of these dependencies. For our fiducial model we find 
  that for a particle mass of 3.3 keV (the 2$\sigma$ lower limit found by
  Viel et al. from a recent analysis of the Lyman-$\alpha$ forest) the Milky Way halo 
  mass is required to be $> 1.4 \times 10^{12}$ \msun. For this same fiducial 
  model, we also find that all WDM particle masses are ruled
  out (at 95\% confidence) if the halo of the Milky Way has a mass
  smaller than $1.1 \times 10^{12}$ \msun, while if the mass of the
  Galactic halo is greater than 1.8 $\times 10^{12}$ \msun,
  only WDM particle masses larger than 2 keV are allowed.
\end{abstract}


\section{Introduction}\label{Introduction}

The nature of the dark matter that makes up most of the matter content
of the Universe is still unknown. There are several particle
candidates which could potentially serve as the dark matter. The
prototype is generically known as a ``weakly interacting massive
particle'', or WIMP, which could be
    the lightest supersymmetric particle, and behaves as cold dark
    matter (CDM; see \citealp{frenkwhite2012} for a review). These
    particles have the property that they acquire negligible thermal
    velocities at early times, giving rise to a power spectrum of
    inflationary density perturbations at recombination that has power
    on all scales; this results in the well-known hierarchical build
    up of cosmic structure. 

    But there are many other candidates which
    are also well-motivated from particle physics. A class of them
    behave as warm dark matter (WDM). These particles acquire
    significant thermal velocities at early times and free-stream
    out of small wavelength perturbations creating a cutoff in the
    linear power spectrum at a wavelength that varies
    roughly inversely with the particle mass. In this case, structure
    formation on scales much larger than the cutoff wavelength proceeds 
    in a very similar way to the CDM case, but the evolution on smaller scales 
    is very different. Good examples of WDM candidates
    are the sterile neutrino (e.g. \citealp{dodelsonwidrow1994, shifuller1999, 
    asakashaposhnikov2005}; see
    \citealp{kusenko2009} for a review), or the
    gravitino (the supersymmetric partner of the graviton; e.g.
    \citealp{pagelsprimack1982, moroi1993, gorbunov2008}). These particles could
    have a mass in the keV range, giving rise to a cutoff in the power
    spectrum on the mass scale corresponding to a dwarf galaxy. A
    mixture of cold and warm dark matter is also possible, for example
    if there is a population of resonantly produced sterile neutrinos
    (\citealp{boyarsky2009}).

    Extensive efforts are underway to detect cold dark matter particles
    either directly in the laboratory, indirectly through annihilation
    products of Majorana particles or at the Large Hadron Collider (see \citealp{strigari2012}
    for a review). None of these searches have produced conclusive
    evidence. While we await developments on the experimental
    front, important conclusions regarding the identity of the dark
    matter may be obtained by confronting predictions for the growth
    of cosmic structure with astronomical data. The key scales to
    distinguish CDM from WDM candidates are subgalactic scales, where
    the effects of the cutoff in the WDM power spectrum are imprinted.
    Furthermore, since the cutoff wavelength depends on the particle
    mass, this approach leads to constraints on the WDM particle mass,
    \mwdm. At high redshift, the relevant scales are only mildly
    non-linear and so calculating the evolution of dark matter, and
    even gas, is relatively straightforward. Using high-resolution
    hydrodynamical simulations to interpret the small-scale clumpiness
    of the Lyman-$\alpha$ flux power spectrum measured from
    high-resolution spectra of 25 $z>4$ quasars, \cite{viel2013}
    have set a lower limit of \mwdm $ \geq 3.3$~keV ($2\sigma$) for
    (thermally produced) warm dark matter particles.

    At the present day, the relevant scales are strongly non-linear
    and so N-body cosmological simulations (or analytical methods
    calibrated on them) are required to predict the evolution of the
    dark matter. The main differences between CDM and WDM are in the
    mass functions and internal structure of halos and subhalos of
    subgalactic mass. For CDM these mass functions increase steeply
    with decreasing mass (e.g. \citealp{Jenkins2001, Tinker2008, 
    Diemand2007, Springel2008}). For WDM the
    abundance of subgalactic mass halos and subhalos is much lower, 
    and has a cutoff at small masses which scales inversely with \mwdm
    \citep{colin2000, bode2001,schneider2012, lovell2013}. In CDM, halos
    and subhalos have cuspy ``NFW'' dark matter density profiles
    (\citealp{Navarro1996, Navarro1997, Springel2008}). In
    WDM cores form but these are much too small to be astrophysically
    relevant (\citealp{maccio2012, shao2013}). In fact, over
    the relevant radial range, the profiles are also cuspy but have
    lower concentration than CDM halos or subhalos of the same mass. The
    central concentration, which reflects the formation time of the
    halo, decreases with decreasing \mwdm (\citealp{avila-reese2001, 
    lovell2012, lovell2013, schneider2012}).

    The differences between CDM and WDM halos and, in the latter case
    the dependence of halo properties on \mwdm, suggest a number of
    astrophysical tests on subgalactic scales that might distinguish
    between the two types of dark matter or set constraints on \mwdm.
    One, based on the different degrees of central dark matter
    concentration between CDM and WDM subhalos, takes advantage of
    recent kinematical data for Milky Way satellites which provide
    information about the distribution of dark
    matter within them. This test is related to the so-called ``too
    big to fail'' problem in CDM: an apparent discrepancy between the
    central dark matter concentration inferred for the brightest dwarf
    spheroidal satellites of the Milky Way and the most massive
    subhalos found in CDM N-body simulations (\citealp{boylan-kolchin2011, 
    boylan-kolchin2012}) and in some gasdynamic simulations that follow 
    the baryonic component of the galaxy including its satellites (\citealp{parry2012}). 
    \cite{lovell2012} showed that the ``too big to fail''
    problem does not exist in simulations of WDM halos with \mwdm$ =
    1.1$~keV\footnote{Some values of \mwdm quoted here differ slightly from those 
    quoted in the original paper, to make them consistent with \citet{viel2005}.}, and 
    \cite{lovell2013} showed that the Milky Way
    satellite data are not sufficiently precise to set an interesting
    upper limit on \mwdm using this test. Even in the case of CDM, the
    ``too big to fail'' problem disappears if the mass of
    the Milky Way halo is less than about $1.5 \times 10^{12} $ \msun
    (\citealp{jwang2012,purcellzentner2012}).

    The second test is based on the different number of subhalos
    predicted to survive in CDM and WDM galactic halos. In the case
    of CDM there are many more subhalos within galactic halos than
    there are observed satellites in the Milky Way, a discrepancy
    often - and incorrectly - dubbed ``the satellite problem in CDM.''
    In fact, it has been known for many years that inevitable feedback
    processes, particularly the early reionization of gas by the first
    stars and winds generated by supernovae, prevent visible galaxies
    from forming in the vast majority of the small subhalos that
    survive inside CDM halos (\citealp{bullock2000, benson2002, 
    somerville2002}). 

    A ``satellite problem,'' however, could exist in
    WDM because if \mwdm is too small, then there will be too few
    surviving substructures to account for the observed number of
    satellites. A limited version of this test was recently applied to
    surviving dark matter subhalos in
        high-resolution N-body simulations of WDM galactic halos by 
        \cite{polisenskyricotti2011}, who found a 
        limit of \mwdm $>2.3$ keV, and by \cite{lovell2013} who
        found a conservative lower limit of
        \mwdm$>1.1$~keV. In this paper we develop this theme further,
        however we apply the test {\it not} to dark matter subhalos but to
        visible
        satellites. This requires following the process of galaxy formation in galactic WDM
    halos, which allows a more direct comparison with observations of
    the Milky Way satellites and leads to stronger limits on
    \mwdm. Since the number of
    surviving subhalos scales with the parent halo mass (\citealp{gao2004}),
    these limits will depend on the mass of the Milky Way halo.
    Unfortunately, this mass is still very uncertain, with estimates
    ranging from about $8 \times 10^{11}$ to $2.5 \times 10^{12} $
    \msun (e.g. \citealp{xue2008, liwhite2008, qiguo2010,
    deason2012, rashkov2013, piffl2013}). 

    In this study we use the Durham semi-analytic model of galaxy
    formation, \galform, to follow galaxy formation in WDM models
    with different values of \mwdm. 
    \cite{nierenberg2013} used a different semi-analytic 
    model to study the redshift evolution of satellite luminosity functions for hosts 
    of different masses, finding that compared to CDM, a \mwdm $=0.75$ keV 
    particle captured better the observed evolution. 
    \cite{macciofontanot2010} 
    also used a semi-analytic model, applied to N-body simulations
    of galactic halos of mass $1.22 \times 10^{12}$~\msun to set a lower
    limit of \mwdm$> 1$~keV. This limit, however, is only valid for halos 
    of this particular mass.
    Here, we use a version of \galform in which galaxy merger trees are computed 
    using Monte Carlo techniques (calibrated on WDM
    N-body simulations). In this way,  we are able to explore models with 
    a wide range of halo masses and thus set limits on
    \mwdm for different values of the, as yet poorly known, Milky Way
    halo mass. Another important advantage of our method is that it
    does not suffer from the problem of spurious halo
    fragmentation which is present in, and complicates the
    interpretation of, high resolution N-body simulations of WDM
    models (\citealp{bode2001, wangwhite2007, lovell2013};
    but see also \citealp{angulo2013}).

    Not surprisingly, only a very minor adjustment to the galaxy
    formation model in CDM is required in WDM to obtain
    a good match to a variety of observed properties of the local 
    galaxy population, such as galaxy luminosity functions in various
    passbands. We then apply this model to derive the expected
    luminosity function of satellites of galaxies like the Milky Way
    and thus set strong constraints on the value of \mwdm as
    a function of the Milky Way halo mass. 

    The rest of this paper is organized as follows:  in Section
    \ref{Models} we introduce our methodology, including the
    computation of the fluctuation power spectrum, the construction of
    merger trees, and the adaptation of our semi-analytic
    model, \galform, to WDM. In Section~\ref{SLFs} we predict
    satellite luminosity functions in galactic halos of different mass
    as a function of \mwdm. In Section~\ref{mX_mH_Relation} we
    discuss the
    range of particle masses that are ruled out based upon various
    estimates of the Milky Way halo mass. A brief discussion of this limit in
    the context of other independent WDM constraints is presented, 
    along with our conclusions, in Section \ref{Conclusions}.

\section{Methods}\label{Models}

\subsection{The warm dark matter linear power spectrum}\label{Warm_Dark_Matter}

In the case where the warm dark matter consists of thermal relics, the
suppression of small-scale power in the linear power spectrum,
$P_{\mathrm{WDM}}$, can be conveniently parametrized by reference to
the CDM power spectrum, $P_{\mathrm{CDM}}$. The WDM transfer function
is then given by,

\begin{equation}
T(k) =  
\Big{[} \frac{P_{\mathrm{WDM}}}  {P_{\mathrm{CDM}}}  \Big{]} ^ {1/2} = 
[ 1 + ( {\alpha}k ) ^{2 \nu} ] ^ {-5 / \nu}
\label{eqn:T(k)}
\end{equation}
\citep{bode2001}. Here, $k$ is the wavenumber and following 
\cite{viel2005} we take the constant $\nu =
1.12$; the parameter $\alpha$ can be related to the mass of the
particle, \mwdm by 

\begin{equation}\label{alphaViel}
\begin{split}
\alpha = 0.049
\Big{(}   \frac{\Omega_\mathrm{WDM}} {0.25}  \Big{)} ^{0.11}     
\Big{(}    \frac{h}{0.7}  \Big{)} ^{1.22}   
\Big{(}   \frac{\mathrm{keV}}{m_{\mathrm{WDM}}}   \Big{)}  ^{1.11} 
h^{-1} \mathrm{Mpc}
\end{split}
\end{equation}
\citep{viel2005}, in terms of the matter density parameter, $\Omega_\mathrm{WDM}$, 
and Hubble parameter, $h=\mathrm{H}_0$ / (100 km s$^{-1}$ Mpc$^{-1}$).

\begin{figure} \begin{center}
    \includegraphics[width=0.46\textwidth]{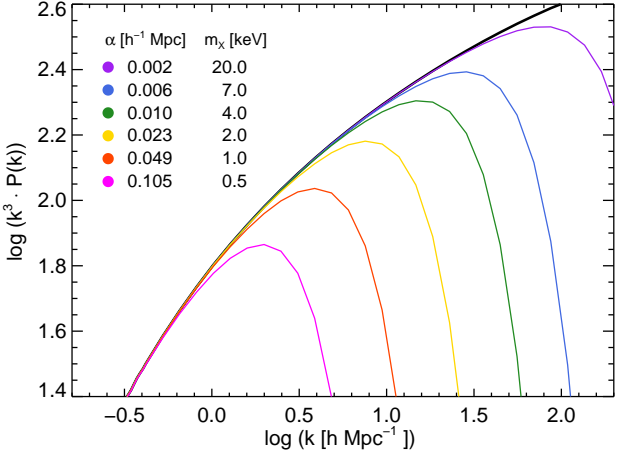}
    \caption{Linear power spectra (in arbitrary units) for warm and cold dark matter
      models. The thick black line shows CDM and the coloured lines
      various WDM models, labelled by their thermal relic
      mass and corresponding value of the damping scale, $\alpha$, in
      the legend.}
\label{WDMPowerSpectra}
\end{center}
\end{figure}

In the case where the WDM particle is a non-resonantly produced
sterile neutrino, its mass $m_{\mathrm{sterile}}$, can be related to
the mass of the equivalent thermal relic, \mwdm, by requiring that the
shape of the transfer function, $T(k)$, be similar in the two cases.
\cite{viel2005} give

\begin{equation}\label{Viel_conversion}
m_{\mathrm{sterile}} = 4.43   \Big{(}   \frac{m_\mathrm{WDM}}{\mathrm{keV}}  \Big{)}   ^{4/3}        \Big{(}   \frac{0.25(0.7)^2}{\Omega_{\mathrm{WDM}} h^2}  \Big{)}  ^{1/3}  \mathrm{ keV}.
\end{equation}
This conversion depends on the specific particle production mechanism
\citep[for a review see][]{kusenko2009}; in the rest of this paper we
will refer only to the thermal relic mass, \mwdm, unless stated
otherwise. We consider particles with masses, \mwdm, ranging from
0.5~keV to 20~keV.  Fig. \ref{WDMPowerSpectra} shows the linear power
spectra for six of the 11 WDM models we have investigated, as well as
for CDM.

We adopt values for the cosmological parameters that are consistent
with the WMAP7 results (\citealp{komatsu2011}): $\Omega_\mathrm{m}=
0.272$, $\Omega_\mathrm{b} = 0.0455$, $\Omega_\Lambda=0.728$,
$h=0.704$, $\sigma_8=0.81$, $n=0.96$. Two hundred merger trees were
generated for each main halo mass and for each WDM particle mass.

\subsection{Galaxy formation models}\label{Galaxy_Formation}

We calculate the properties of the galaxy population in our WDM models
using the Durham semi-analytic galaxy formation model, \galform (e.g.
\citealp{cole2000, benson2003, bower2006}). Rather than applying it to
merger trees obtained from an N-body simulation, we instead construct
Monte Carlo merger trees using the Extended Press-Schechter (EPS)
formalism (\citealp{pressschechter1974, bond1991, bower1991, laceycole1993,
  parkinson2008}) to generate conditional mass functions for halos of
a given mass. The standard formulation of the EPS formalism (in which
the density field is filtered with a top hat in real space) is not
applicable in the presence of a cutoff in the power spectrum. Instead,
using a sharp filter in $k$-space produces a halo mass function in
good agreement with the results of N-body simulations. We adopt this
prescription which is justified and described in detail in
\cite{benson2013}. A similar procedure was adopted by Schneider et al.
(2013) but other authors, such as \cite{smithmarkovic2011} and
\cite{menci2012}, have used a top hat filter in real space and then
multiplied the resulting mass function by an {\it{ad hoc}} suppression
factor. We do not apply the correction for finite phase-space density
derived by \cite{benson2013} because the effect of thermal velocities
is negligible in the models we consider (\citealp{maccio2012,
  shao2013}). Halo concentrations were set according to the NFW
prescription \citep{Navarro1996, Navarro1997}, as described in
\cite{cole2000}, thus explicitly taking into account the later
formation epoch of WDM halos compared to CDM halos of the same mass.
These concentrations are broadly in agreement with the WDM simulations
of \cite{schneider2012}.

We use the latest version of \galform (Lacey et al. 2013, in prep.)
which includes several improvements to the model described by
\cite{bower2006}.  The standard \galform model is tuned to fit a set
of observed properties of the local galaxy population assuming CDM.
Thus, an adjustment is required in the WDM case. On scales larger than
dwarf galaxies at $z=0$ there is little difference between WDM and CDM
models. On smaller scales, the most important processes that influence
galaxy formation are the feedback effects produced by the early
reionization of the intergalactic medium and supernova feedback.

In \galform, reionization is modelled by assuming that no gas is able
to cool in galaxies of circular velocity less than \vcut at
redshifts less than \zcut. For CDM, the values 
\vcut$=30$~km~s$^{-1}$ and \zcut$=10$ result in a good
    approximation to more advanced treatments of reionization 
    (\citealp{okamoto2008, font2011}). Supernovae feedback, on the other
    hand, is controlled by the parameter $\beta$, the ratio of the
    rate at which gas is ejected from the galaxy to the star formation
    rate. This ratio is assumed to depend on the circular velocity of
    the disc, $v_{\mathrm{circ}}$, as:

\begin{equation}
\beta =  \Big{(}  \frac{v_{\mathrm{circ}}} {v_{\mathrm{hot}}}  \Big{)}
^{-{\alpha_{\mathrm{hot}}}}, 
\label{eqn:feedback}
\end{equation}
where $v_{\mathrm{hot}}$ and \ahot are adjustable parameters fixed
primarily by the requirement that the model should match the local
$b_J$- and $K$-band galaxy luminosity functions. In the Lacey et al.
model, these parameters take on the values
$v_{\mathrm{hot}}=300$~km~s$^{-1} $ and \ahot= 3.2. Since
$v_{\mathrm{circ}}$ depends on the concentration of the host halo,
which is lower for a WDM halo than for a CDM halo of the same mass
(\citealp{lovell2012}), we expect that a small adjustment to the
parameters in eqn.~\ref{eqn:feedback} will be required to preserve the
good match to the local luminosity functions. 

\begin{figure} 
\begin{center} 
\includegraphics[width=0.46\textwidth]{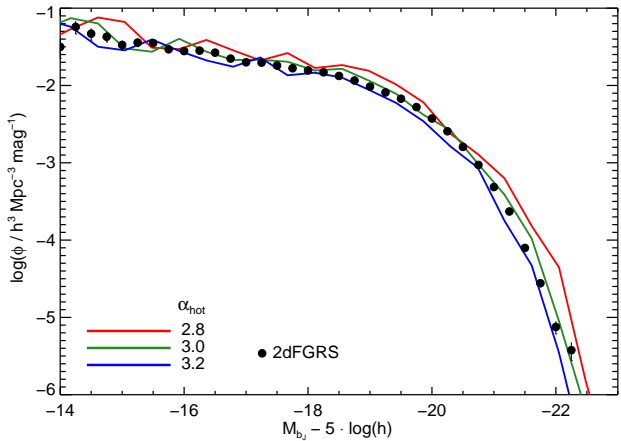} 
\caption{The $b_{J}$-band local galaxy luminosity function for \mwdm
  $=3$ keV compared to the 2dFGRS determination (indicated by
  circles). Coloured curves show the effect of varying \ahot, as shown
  in the legend}
\label{alphahot_plots}
\end{center}
\end{figure}

Fig.~\ref{alphahot_plots} shows the $b_{J}$-band field galaxy
luminosity function for different values of \ahot for the case of a 2
keV particle. Here, \vcut and \zcut are set to the CDM values. (The
reionization model mostly affects galaxies fainter than those included
in estimates of the field luminosity function). The figure shows that
only a small change in the value of $\alpha_{\mathrm{hot}}$ is
required to achieve as good a fit to the measured $b_J$-band
luminosity function as in the CDM case. The best fit for \mwdm$=3$~keV
is obtained for $\alpha_{\mathrm{hot}} \sim 3.0$ (green line; assuming the
same value of $v_{\mathrm{hot}}=300$~km~s$^{-1} $ as in CDM). In
general, we find that the local galaxy luminosity function in WDM
models is well reproduced for a wide range of values of \mwdm by
setting,

\begin{equation}
\alpha_{\mathrm{hot}} (m_{\mathrm{WDM}})= 3.2 - 0.3 \left(\frac {m_{\mathrm{WDM}}}  {\mathrm{keV}}\right) ^{-1}
\label{ahot_relation}
\end{equation}
(keeping the same values of $v_{\mathrm{hot}}$ and of \vcut and \zcut
as above). This adjustment also results in acceptable matches to the
$K$-band luminosity function, Tully-Fisher relation, size distribution
and other observables. However, we find that for \mwdm$<1.5$~keV, we
cannot obtain acceptable models using eqn.~\ref{ahot_relation}.
\cite{kang2013} also found that it was not possible to find a consistent
model of galaxy formation for such low mass WDM particles. Since these
masses are, in any case, ruled out by observations of the
Lyman-$\alpha$ forest, we restrict the rest of this analysis to the 9
models with particle masses larger than 1.5 keV.

In Section~\ref{VariousModels} we vary the adjustable parameters in
our models of reionization and supernovae feedback to assess how they
affect our inferred lower limits on the WDM particle mass. Throughout
the remainder of this paper, we will refer to the model described here
as the `fiducial' model.

\section{Satellite Luminosity Functions}\label{SLFs}

We now consider satellite systems, firstly those predicted by \galform to exist in 
halos of mass similar to that of the Milky Way's, and then the Milky Way's own 
system. We then describe the method we have adopted to compare the two.

\subsection{The predicted satellite population}\label{sats_predictions}

We use the models described in Section~\ref{Warm_Dark_Matter} with
final halo masses ranging from $5 \times 10^{10}$ \msun to $1 \times
10^{13}$ \msun, a significantly wider range than that covered by
recent estimates of the Milky Way's halo mass. The mass resolution of
the merger trees is set to $1 \times 10^6$ \msun, which is below the
free-streaming scale of our WDM models.

\begin{figure} \begin{center}
    \includegraphics[width=0.48\textwidth]{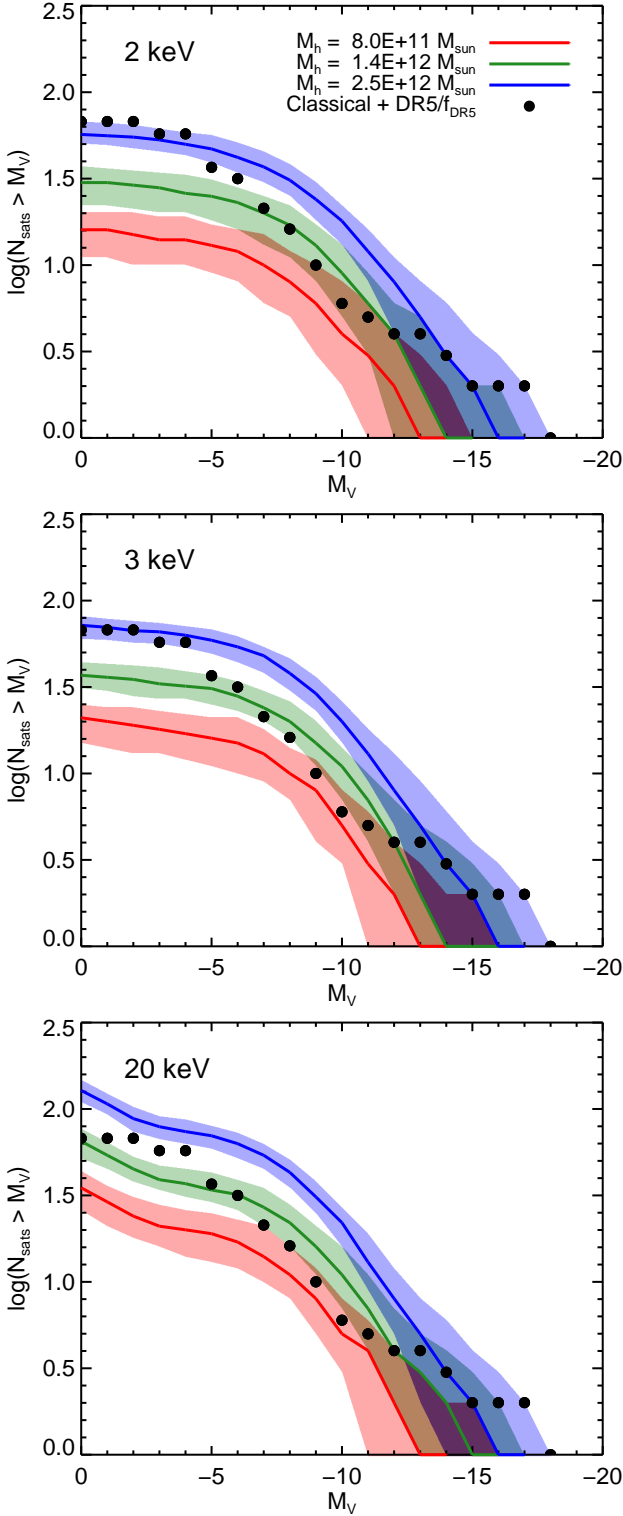}
    \caption{Satellite galaxy luminosity functions predicted by our
      fiducial semi-analytic model in galactic halos of different
      mass, for WDM particle masses, \mwdm, of 2~keV, 3~keV and
      20~keV, as indicated in the legend. The different coloured
      curves correspond to different host halo mass. The solid line in
      each case is the median cumulative V-band satellite luminosity
      function and the edges of each band indicate the 10th and 90th
      percentiles. For reference, the luminosity function of the 11
      observed classical satellites, plus the DR5 satellites (scaled
      for sky coverage assuming an isotropic distribution) is
      indicated by the black dots.}
\label{SLFs4mX}
\end{center}
\end{figure}

Fig.~\ref{SLFs4mX} shows the predicted cumulative V-band satellite
luminosity functions for several examples. The three panels show
results for \mwdm = 2, 3 and 20~keV and, within each panel, the effect
of increasing the host halo mass from $8 \times 10^{11} $ \msun to
$2.5 \times 10^{12} $ \msun is demonstrated. Increasing the host halo
mass increases the number of satellites at all luminosities, and
increasing the WDM particle mass increases the number of satellites
particularly at fainter magnitudes. The number of bright satellites
($\rm M_{\rm V} \lsim -12$) is insensitive to \mwdm because these
satellites form in halos with mass above the cutoff scale in
the WDM power spectrum.

\subsection{The observed satellite population}\label{Obs_Data}

To determine whether a model produces a satisfactory number of
satellites we make use of observations of the satellites around the
Milky Way. While there have been recent censuses of satellites around
galaxies outside the Local Group (e.g.
\citealp{quanguo2011,lares2011,liu2011,wwang2012,
  strigariwechsler2012}) these tend to be limited to the brightest
few. Many faint satellites have been observed around M31 (e.g.
\citealp{martin2006, martin2009, martin2013, ibata2007,
  mcconnachie2009}), but in this analysis we limit ourselves to
studies of the population in our own galaxy.

There are eleven bright satellite galaxies around the Milky Way which
were discovered in the previous century; these are dubbed the
``classical satellites''. In more recent years, the Sloan Digital Sky
Survey (e.g. \citealp{adelman-mccarthy2007}) has revealed a number of
fainter satellite galaxies. For this analysis we focus on 11
additional satellites found in the SDSS Data Release~5 (DR5) (see
summary in \citealp{tollerud2008}), not double counting any classical
satellites. This survey covers a fraction $f = 0.194$ of the sky, which 
is roughly 8000 square degrees, to
a depth of around 22.2 in the $g$- and $r$- bands. We refer to these
satellites here as the ``DR5 satellites''.

It is likely that there are yet more satellites in the DR5 region
which have not been detected due to their faintness; at 260 kpc the
survey is only complete to $M_{V} \approx -6$ (\citealp{koposov2008}).
Attempts to correct for the detection limits of the survey by assuming
a given radial profile of the satellites predict a total satellite
population of hundreds (e.g. \citealp{koposov2008, tollerud2008}).

\subsection{Assessing model population likelihoods}\label{Likelihoods_of_Models}

For the purposes of comparing our model predictions with satellite
galaxy data, we will consider only those satellites brighter than
$M_{V}$ = -2, which is fainter than the magnitude of all the DR5
satellites.
Since \galform only makes predictions for satellites which lie within
the virial radius of the host halo, we limit our analysis of the real
Milky Way satellites to those with a galactocentric distance less than
the virial radius of a particular halo in the semi-analytic
calculation. Here, the virial radius is defined as the boundary of the
region enclosing an overdensity, $\Delta$, with respect to the
critical density, where, for the spherical collapse model, $\Delta
\approx 93$ (\citealp{eke1996}).

In order to estimate the total number of satellites brighter than
$M_{V} = -2$ that we would expect around the Milky Way, it is
necessary to make some assumptions about the underlying distribution
since it is not fully sampled. Firstly, we make the assumption that
all the `classical' satellites (those with apparent magnitudes
brighter than $M_{V} \approx -8.5$) have been observed. This is
probable, although our results would not change significantly even if
one or two remained undetected behind the Milky Way disk.

Next, we assume that the underlying distribution of satellites is
isotropic, so that the DR5 represents a geometrically unbiased
sampling. This may be unrealistic because the eleven classical
satellites of the Milky Way are known to lie in a `pancake' structure
oriented approximately perpendicular to the plane of the Milky Way
disk (\citealp{lynden-bell1976, lynden-bell1982, majewski1994, libeskind2005}). A
large region of the DR5 footprint intersects this plane; if as yet
undetected satellites also tend to lie in this disk, then the DR5
would provide a biased sampling of the true satellite population,
leading us to overpredict the number of satellites that are
necessary to match the data. This would have the effect of weakening
our lower limit on \mwdm. However, cosmological N-body simulations
show that the preferentially flattened satellite distributions are
restricted to the brightest satellites, and that as fainter and
fainter populations are considered, their distribution tends to become
increasingly isotropic (\citealp{jwang2013}).

Finally, we make the extremely conservative assumption that every
satellite in the DR5 footprint area has been detected, so that no more
faint satellites are lurking below the detection threshold. Given the
survey's radial completeness limits, this is unrealistic. This
assumption works in the sense of making our inferred lower limits on
\mwdm conservative. If future or current surveys, such as Pan-STARRS,
were to reveal even more faint satellites, our lower mass limits would
become correspondingly stronger.

To quantify whether the model satellite population is compatible with
the MW data, we require that the model should produce at least as many
satellites with $M_{V} < -2$ as are known to exist in the Milky Way.
To find the likelihood of each model given the data, we calculate the
probability that the predicted satellite population includes at least
as many members falling within a region the size of the DR5 footprint,
i.e. covering a fraction of the sky, $f = 0.194$, as the DR5 survey
itself, which contains $n_{\mathrm{DR5}}$ satellites\footnote{The
  value of $n_{\mathrm{DR5}}$ used will depend upon the virial radius
  of the halo we compare to.}.

First, we define the number of classical Milky Way satellites (again
within the virial radius of the model halo) to be
$n_{\mathrm{class}}$. This number is subtracted from the total number
of predicted satellites, $n_{\mathrm{galform}}$, to prevent
double-counting in the DR5 region,

\begin{equation}
n_{\mathrm{pred}} = n_{\mathrm{galform}} - n_{\mathrm{class}}
\end{equation}
Then, for this remaining population of $n_{\mathrm{pred}}$ satellites,
we must find the likelihood that they are distributed such that at
least as many satellites as are observed in DR5 fall in a region
covering a fraction $f$ of the total sky area. We find the
probability, $P$, that a number between $n_{\mathrm{DR5}}$ and
$n_{\mathrm{pred}}$ satellites lie in this region by assuming that a
given satellite is equally likely to be found anywhere on the sky.
Hence, $P$ can be calculated from a binomial distribution,

\begin{equation}\label{sum_binomial}
P = 
\sum_{k=n_{\mathrm{DR5}}}^{k=n_{\mathrm{pred}}}  
\Big{(}\frac{n_{\mathrm{pred}}!} {k! ( n_{\mathrm{pred}}-k)!} \Big{)}     
\cdot     f^k     
\cdot   (1-f)^{n_{\mathrm{pred}}-k}
\end{equation}
Eqn.~\ref{sum_binomial} gives the probability that any given
realization of a halo merger tree, for a particular value of \mwdm,
within a given host halo mass, $M_h$, has produced enough satellites
to be compatible with the Milky Way data. Since we have generated 200
merger trees for each WDM model at a given host halo mass, we take the
average of the probabilities, $P$, computed for each individual host
halo using eqn.~\ref{sum_binomial}.

If $\langle P \rangle$ is smaller than 0.05, we conclude that this
model predicts too little substructure to account for the
observations. Conversely, for each WDM particle mass, \mwdm, we find
the minimum host halo mass, $M_{h}$, for which $\langle P \rangle$ is
larger than 5\%. This value of \mwdm is therefore the limiting mass
that cannot be excluded at 95\% confidence.

\section{Results: limits on the WDM particle mass}\label{mX_mH_Relation}

In this section we present the constraints\footnote{These data can be 
accessed by contacting the lead author.} on the warm dark particle mass 
that follow from comparing our predictions for the satellite luminosity functions 
with the Milky Way data. We also discuss how our limits can be affected by 
uncertainties in our modelling of galaxy formation. 

\subsection{Fiducial model}\label{fiducial}

\begin{figure*} \begin{center}
    \includegraphics[width=1.0\textwidth]{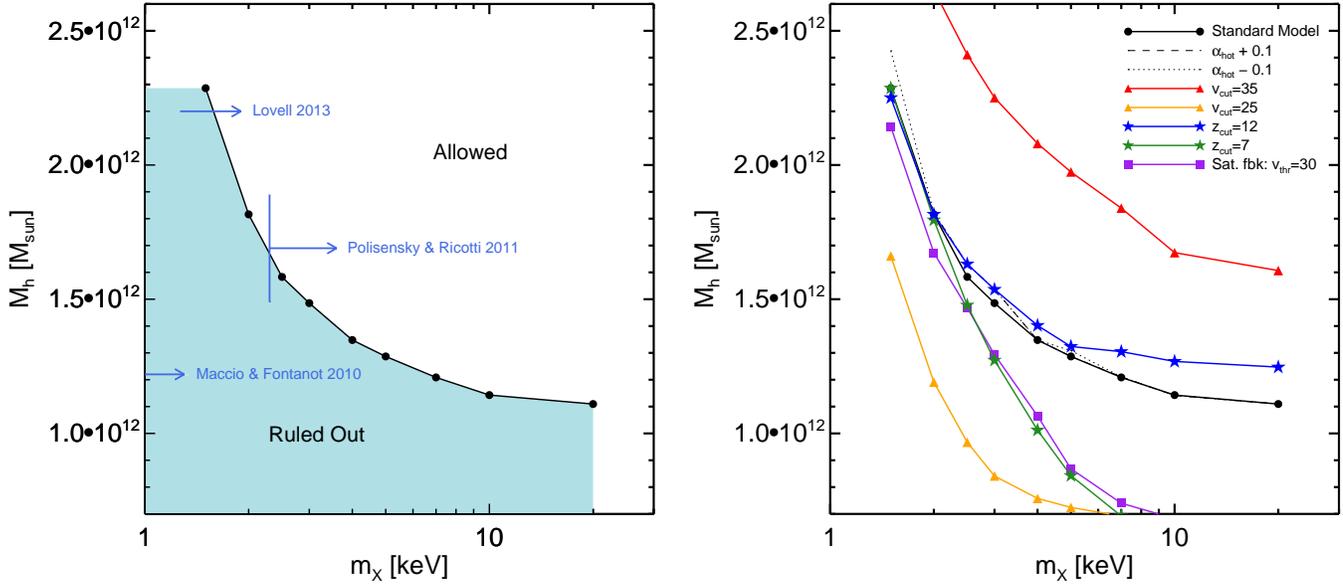}
    \caption{Left: exclusion diagram for thermal WDM particle masses,
      \mwdm, as a function of the Milky Way dark matter halo mass,
      $M_h$; the shaded region is excluded. The lower limits reported by other authors, 
      as well as the host halo masses they considered, are indicated by the arrows.
      Right: sensitivity of our
      constraints to variations in the parameters of our galaxy
      formation model; the lines show the envelope of the exclusion
      region.}
\label{mX_Mh}
\end{center}
\end{figure*}

The constraints on the WDM particle mass as a function of host halo
mass set by the method described in
Section~\ref{Likelihoods_of_Models} are shown in the exclusion diagram
of Fig.~\ref{mX_Mh}(a). Each point in the plot gives the smallest
Galactic halo mass that has at least a $5\%$ chance of hosting enough
satellites to account for the observed number. Conversely, for a given
Galactic halo mass, the minimum allowed WDM particle mass can be read
off the $x$-axis. The shaded region shows the parameter space
that is excluded. For example, if the Milky Way were found to have a
mass of $1.5 \times 10^{12}$~\msun, then the thermal relic dark matter
particle must be more massive than 3~keV. The envelope of the
exclusion region asymptotes to a value of $1.1 \times 10^{12}$~\msun.
Thus, for Milky Way halo masses below this value, all WDM
particle masses are ruled out at 95\% confidence by our model.

An accurate measurement of the Milky Way's halo mass, $M_{\mathrm{h}}$, could, in
principle, rule out all astrophysically interesting thermally-produced
WDM particles. Unfortunately, this measurement is difficult and
subject to systematic uncertainties. Several methods have been used to
estimate $M_\mathrm{h}$. (The values quoted below refer to different
definitions of virial mass assuming different values of the limiting
density contrast, $\Delta$, as indicated by the subscript,
$M_{\Delta}$). A traditional one is the timing argument of \cite{kahnwoltjer1959}
 which employs the dynamics of the Local Group to
estimate its mass. Calibrating this method with CDM N-body
simulations, \cite{liwhite2008} find $M_{200} \sim 2.43 \times 10^{12}
$~\msun, with a lower limit of $M_{200} = 8.0 \times 10^{11}$~\msun at
95\% confidence. A rather different method is based on matching the
abundance of galaxies ranked by stellar mass to the abundance of dark
matter halos ranked by mass in a large CDM N-body simulation. This
technique gives upper and lower 10\% confidence limits of $8 \times 
10^{11} < M_{\mathrm{200}} < 4.7 \times 10^{12} $~\msun
(\citealp{qiguo2010}).

A third class of methods relies on the kinematics of tracer stars in
the stellar halo to constrain the potential out to large distances.
Using positions and line-of-sight velocities for 240 halo stars,
\cite{battaglia2005} find $6 \times 10^{11} <
M_{\mathrm{100}} < 3 \times 10^{12}$~\msun, depending on assumptions
about the halo profile; using 2000 BHB stars out to 60 kpc,
interpreted with the aid of simulations, \cite{xue2008} find
$8 \times 10^{11} < M_{\mathrm{102}} < 1.3 \times 10^{12}$~\msun.
Using a variety of tracers, \cite{deason2012} find the mass within 150 kpc
to be between $5 \times 10^{11}$ \msun and $1 \times 10^{12}$ \msun.
Most recently, \cite{piffl2013} used a large sample of
stars from the RAVE survey in conjunction with cosmological
simulations to find 
$1.3 \times 10^{12} < M_{\mathrm{200}} < 1.8 \times 10^{12}$~\msun.

\subsection{Sensitivity to galaxy formation model parameters}
\label{VariousModels}

Given an assumption about the nature of the dark matter, the abundance
of galactic satellites depends primarily on two key astrophysical
processes: the reionization of hydrogen after recombination and
feedback from supernovae explosions. The epoch during which the
Universe became reionized is constrained by temperature anisotropies
in the microwave background and their polarization to lie in the range
$8 \lesssim z_{\mathrm{re}} \lesssim 14$ (\citealp{plankcollabI2013}).
Photoheating raises the entropy of the gas and suppresses cooling into
halos of low virial temperature.

In \galform reionization is modelled by assuming that no gas cools in
halos of circular velocity smaller than \vcut at redshifts lower than
\zcut. This simple prescription has been shown to be a good
approximation to a more detailed semi-analytic model of reionization
(\citealp{benson2002}) and to full gasdynamic simulations (\citealp{okamoto2008}).
In our fiducial model, the parameters take the values \vcut
$=30$~km s$^{-1}$ and \zcut$=10$. The simulations of \cite{okamoto2008}
suggest that \vcut is around 25~km~s$^{-1}$, but \cite{font2011}
   conclude that a value of \vcut $=34$~km~s$^{-1}$ is
    required to match the results of the detailed semi-analytical calculation 
    of the effects of reionization given by \cite{benson2002}.  
    We explore the effect of varying both \vcut and \zcut within these bounds.

    Supernova feedback is still poorly understood. In \galform, this
    process is modelled in terms of a simple parametrized power-law of
    the disc circular velocity with exponent \ahot
    (eqn.~\ref{eqn:feedback}). As discussed in
    Section~\ref{Galaxy_Formation}, the parameter \ahot is constrained
    -- as a function of \mwdm -- by the strict requirement that the
    model should provide an acceptable fit to the observed local
    $b_J$-band galaxy luminosity function. This is a strong constraint
    which limits any possible variation of \ahot to less than
    ${\pm}0.1$. Our simple parametrization ignores, for example,
    environmental effects (\citealp{lagos2013}) but these are unlikely
    to make a significant difference to our conclusions so we do not
    consider them further. However we do consider a model in which the
    effects of feedback saturate below $v_{\mathrm{circ}}$ = 30~km~s$^{-1}$, 
    similar to what \cite{font2011} argue is required to explain the
    variation of metallicity with luminosity observed in the
    population of Milky Way satellites.

    The effects of varying the galaxy formation model parameters
    (retaining agreement with the local field galaxy luminosity
    function) on our constraints on \mwdm as a function of $M_\mathrm{h}$ are
    shown in Fig. \ref{mX_Mh}(b). Varying \ahot has a very small
    effect; varying \zcut affects, to some extent, the
    limits for WDM particle masses greater than 2-3 keV. The main
    sensitivity is to the parameter \vcut which has a strong
    effect on the number of small halos which are able to form stars. 
    At fixed halo mass, lower values of \vcut weaken the limits on \mwdm 
    whereas larger values strengthen them. The range considered here, 
    $25<v_{\mathrm{cut}}/{\rm km}~\mathrm{s^{-1}} <35$, is realistic according to
    current understanding of the process of reionization. 

\section{Discussion and conclusions}\label{Conclusions}

The cutoff in the linear power spectrum of density fluctuations
produced by the free streaming of warm dark matter particles in the
early universe provides, in principle, the means to search for
evidence of these particles. If the particle mass is in the keV range,
the cutoff occurs on the scale of dwarf galaxies and no primordial
fluctuations are present on smaller scales. Thus, establishing how
smooth the universe is on these scales could reveal the existence of
WDM or, since the cutoff length scales inversely with the particle
mass, set limits on its mass. The traditional method for testing the
smoothness of the density field at early times is to measure the flux
power spectrum of the Lyman-$\alpha$ forest in the spectra of high
redshift quasars. The most recent lower limit on the WDM particle mass
using this method on data at redshifts $z \sim 2$$-$$6$ is that set by
\cite{viel2013}, \mwdm$ \geq 3.3$~keV ($2\sigma$), for thermally
produced warm dark matter particles.
  
A different way to estimate the clumpiness of the matter density field
on small scales, this time at the present day, is to count the number
of substructures embedded in galactic halos. The most direct way to do
this is to count the satellites that survive in such halos but these
are so faint that sufficient numbers can only be found in our own
Milky Way galaxy and M31. Counting the Milky Way satellites thus provides a
test of WDM which is independent from and complementary to the
Lyman-$\alpha$ forest constraint. There are several complications that
need to be taken into account when carrying out this test. Firstly, a
suitable property to characterize the satellite population needs to be
identified. The maximum of the circular velocity curve,
$v_{\mathrm{max}}$, is often used for this purpose, but this quantity
is not directly measurable for the Milky Way's satellites. The
luminosities of satellites, on the other hand, are accurately
measured, but using this as a test of WDM requires the ability to
predict the satellite luminosities and this, in turn, requires
modelling galaxy formation. This is the approach we have adopted in
this paper where we have made use of the semi-analytic model,
\galform. This model has the virtue that it gives a good match to the
field galaxy luminosity function in various bands and has been
extensively tested against a variety of other observational data. The
$v_{\mathrm{max}}$ test was carried out by \cite{polisenskyricotti2011}
and by \cite{lovell2013} but the uncertainty in the
satellites' values of $v_{\mathrm{max}}$ introduces some uncertainty in the limits set.

The second complication is the requirement to understand the
completeness of the satellite sample. The Milky Way has a population
of 11 bright or ``classical'' satellites which is thought to be
complete (although one or two bright satellites could be lurking
behind the Galactic Plane, too small a number to affect our
conclusions) and a population of faint and ultrafaint satellites that
have been discovered in the fifth of the sky surveyed by the SDSS.
While the classical satellites are known to be distributed on the thin
plane, identified by \cite{lynden-bell1976}, it is not known
if the SDSS sample is also anisotropic. Large N-body CDM simulations
suggest that it is only the brightest satellites that lie on a plane
whereas more abundant populations tend to be much less anisotropically
distributed (\citealp{jwang2013}). Here we assume that the spatial
distribution of the Milky Way satellites other than the classical ones
is isotropic. If this assumption were incorrect, we would overestimate
the number of satellites which would cause us to overestimate the
minimum WDM particle mass required to have enough satellites in a halo
of a given mass. The simulations of \cite{jwang2013} suggest that
this effect is unlikely to be large.

The third complication of our method is the difficulty in assessing
possible systematic effects arising from uncertainties in our galaxy
formation model. As we discussed in Section~\ref{VariousModels}, the
main source of uncertainty is our treatment of the inhibiting effect
of the early reionization of the intergalactic medium on the cooling
of gas in small halos. We model this process in a relatively simple
way which, however, has been validated both by realistic semi-analytic
calculations (\citealp{benson2002}) and by full cosmological
hydrodynamic simulations (\citealp{okamoto2008}).
Another uncertainty arises from the fate of satellites prior to merging with 
the central galaxy: we do not currently consider tidal disruption effects in our 
model, meaning that all satellites survive until the point of merging. If tidal 
destruction is an important phenomenon, which may be especially true for 
WDM, then we would expect fewer surviving satellites in our models. This 
would have the net effect of increasing further our lower limits on \mwdm.

Since the number of surviving subhalos is a strong function of the
parent halo mass, our limits on \mwdm depend on the mass of the Milky
Way halo which, unfortunately, is still uncertain to within a factor
of at least a few.  For our fiducial model of galaxy formation, we
find that if the halo mass is less than $1.1 \times 10^{12}$~\msun, then
{\it all} values of \mwdm are ruled out at 95\% confidence for the
case of thermally-produced WDM particles. If, however, the mass of the
halo is greater than $1.3 \times 10^{12}$~\msun, then, at the same
confidence level, all masses greater than \mwdm$=5$~keV are allowed
and if it is greater than $2 \times 10^{12}$~\msun, then all masses
greater than \mwdm = 2~keV are allowed. If the main parameter in our
model of reionization, \vcut, had a value of
$35$~km~$\mathrm{s^{-1}}$, then most (thermal) masses of astrophysical
interest would be ruled out even if the mass of the halo is $2 \times
10^{12}$ \msun, but if this parameter is only 25~km~$\mathrm{s^{-1}}$,
then only masses below \mwdm = 2.5~keV are ruled out for halo masses
less than $1 \times 10^{12}$ \msun. By contrast, using the abundance
of dark matter subhalos as a function of $v_{\mathrm{max}}$,
\cite{lovell2013} were only be able to set a lower limit of \mwdm
$=1.3$~keV for dark matter halos of mass $1.8 \times 10^{12}$ \msun.

Our limits on the WDM particle mass from the abundance of satellites
in the Milky Way are compatible with those set by the Lyman-$\alpha$
forest constraints, except, of course, that they depend on the mass of
the Milky Way halo. The value of the most recent lower limit (\mwdm $
=3.3$~keV) derived from the Lyman-$\alpha$ forest requires the halo
mass to be $M_\mathrm{h} > 1.4 \times 10^{12}$~\msun in order for there to be enough
satellites in the Milky Way.  All these limits apply only to thermally
produced WDM and need not exclude specific warm candidates such as
sterile neutrinos. In this case (and also for other types of WDM),
there could also be additional resonantly produced particles that
could behave as cold dark matter, resulting in a different small scale
behaviour of the linear density power spectrum, depending on the mass
and formation epoch of these particles.

Sterile neutrinos can decay and emit a narrow X-ray line. The absence
of such a line in the X-ray spectra of galaxy clusters can be used to
set an {\it upper} limit to \mwdm but this depends in the sterile
neutrino production mechanism. For example, for non-resonant
production, \cite{abazajian2001} have set an {\it upper} limit of
$m_{\mathrm{sterile}} \lesssim 5$ keV which would correspond to a
thermal mass of $\sim 1$~keV. 

The constraints presented in this study would become much tighter if
the mass of the Milky Way halo could be measured accurately.  While
the recent RAVE results (\citealp{piffl2013}) are encouraging, it is to be
hoped that the forthcoming GAIA satellite mission will allow a better
understanding of the systematic effects that complicate these kinds of
measurements.  In the meantime, gravitational lensing effects such as
the flux ratio anomaly in multiply-lensed quasar images may provide a
direct measurement of the amount of substructure present in galactic
dark matter halos (\citealp{mirandamaccio2007, ddxu2013}). This is a 
powerful method that could, in principle, provide a conclusive test of 
whether the dark matter is cold or warm. 

\section{Acknowledgements}

We thank Mark Lovell for useful discussions.  CSF
acknowledges an ERC Advanced Investigator grant, COSMIWAY. The
calculations performed for this work were carried out on the Cosmology
Machine supercomputer at the Institute for Computational Cosmology,
Durham.  The Cosmology Machine is part of the DiRAC Facility jointly
funded by STFC, the Large Facilities Capital Fund of BIS, and Durham
University. This work was supported in part by the STFC rolling grant ST/F001166/1
to the ICC and by the National Science Foundation under Grant No.
PHYS-1066293.  CSF acknowledges the hospitality of the Aspen Center
for Physics.

\bibliographystyle{mn2e}
\bibliography{Bibliography}

\end{document}